\documentclass[aps,prb,
preprint,
superscriptaddress,
floatfix,
amsmath
]{revtex4-1}

\usepackage{graphicx}
\usepackage{braket}
\usepackage{multirow}

\begin{document}

\title{Intermolecular interactions and substrate effects for an adamantane
  monolayer on the Au(111) surface}

\author{Yuki Sakai}
\affiliation{Department of Physics, Tokyo Institute of Technology, 2-12-1 Oh-okayama, Meguro-ku, Tokyo 152-8551, Japan}
\affiliation{Department of Physics, University of California, Berkeley, California 94720, USA}
\author{Giang D.~Nguyen}
\affiliation{Department of Physics, University of California, Berkeley, California 94720, USA}
\author{Rodrigo B.~Capaz}
\affiliation{Department of Physics, University of California, Berkeley, California 94720, USA}
\affiliation{Instituto de Fisica, Universidade Federal do Rio de Janeiro, Caixa Postal 68528, Rio de Janeiro, RJ 21941-972, Brazil}
\author{Sinisa Coh}
\affiliation{Department of Physics, University of California, Berkeley, California 94720, USA}
\affiliation{Materials Sciences Division, Lawrence Berkeley National Laboratory, Berkeley, California 94720, USA}
\author{Ivan V.~Pechenezhskiy}
\affiliation{Department of Physics, University of California, Berkeley, California 94720, USA}
\affiliation{Materials Sciences Division, Lawrence Berkeley National Laboratory, Berkeley, California 94720, USA}
\author{Xiaoping Hong}
\affiliation{Department of Physics, University of California, Berkeley, California 94720, USA}
\author{Feng Wang}
\affiliation{Department of Physics, University of California, Berkeley, California 94720, USA}
\affiliation{Materials Sciences Division, Lawrence Berkeley National Laboratory, Berkeley, California 94720, USA}
\author{Michael F.~Crommie}
\affiliation{Department of Physics, University of California, Berkeley, California 94720, USA}
\affiliation{Materials Sciences Division, Lawrence Berkeley National Laboratory, Berkeley, California 94720, USA}
\author{Susumu Saito}
\affiliation{Department of Physics, Tokyo Institute of Technology, 2-12-1 Oh-okayama, Meguro-ku, Tokyo 152-8551, Japan}
\author{Steven G.~Louie}
\affiliation{Department of Physics, University of California, Berkeley, California 94720, USA}
\affiliation{Materials Sciences Division, Lawrence Berkeley National Laboratory, Berkeley, California 94720, USA}
\author{Marvin L.~Cohen}
\affiliation{Department of Physics, University of California, Berkeley, California 94720, USA}
\affiliation{Materials Sciences Division, Lawrence Berkeley National Laboratory, Berkeley, California 94720, USA}

\date{\today}

\begin{abstract}

   We study theoretically and experimentally the infrared (IR) spectrum 
   of an adamantane monolayer on a Au(111) surface. 
   Using a new STM-based IR spectroscopy technique (IRSTM) 
   we are able to measure both the nanoscale structure of an adamantane
   monolayer on Au(111) as well as its infrared spectrum, 
   while DFT-based ab initio calculations allow us to interpret the microscopic
   vibrational dynamics revealed by our measurements. We find that the 
   IR spectrum of an adamantane monolayer on Au(111) is substantially 
   modified with respect to the gas-phase IR spectrum. The first 
   modification is caused by the adamantane--adamantane interaction due to 
   monolayer packing and it reduces the IR intensity of the 2912~cm$^{-1}$
   peak (gas phase) by a factor of 3.5. The second modification originates from the 
   adamantane--gold interaction and it increases the IR intensity of
   the 2938~cm$^{-1}$ peak (gas phase) by a factor of 2.6, and reduces its 
   frequency by 276~cm$^{-1}$. We expect that the techniques described here can be 
   used for an independent estimate of substrate effects and 
   intermolecular interactions in other diamondoid molecules, and for 
   other metallic substrates. 
\end{abstract}


\maketitle

\section{Introduction}
\label{sec:introduction}

\begin{figure*}
\includegraphics[width = 16.5cm]{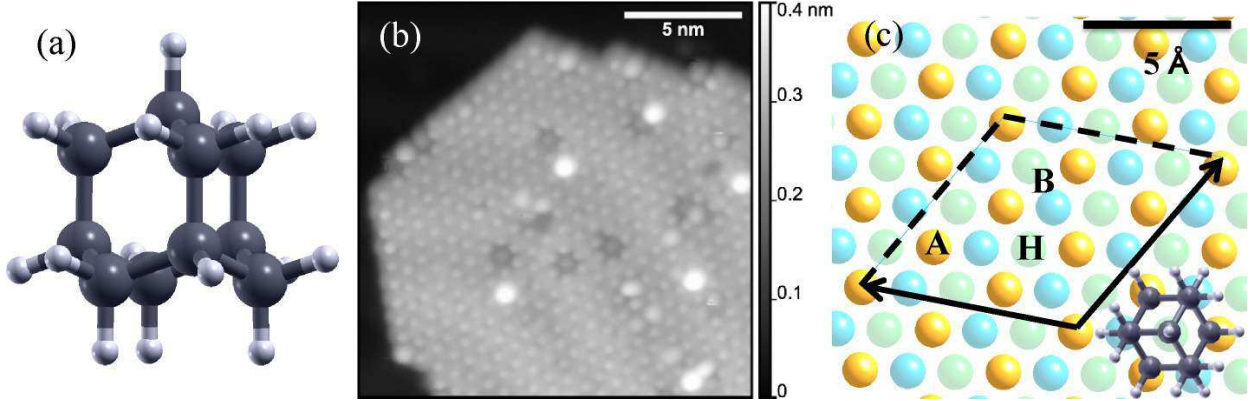}
\caption{\label{geom} (Color online) (a) Model of molecular structure of an
  adamantane molecule.  Hydrogen and carbon atoms are represented by
  white and gray spheres, respectively.  (b) 
  STM topography of a
  self-assembled island of adamantane molecules on a Au(111) surface
  (sample bias $V_{\rm sample} = -1.0$~V, current setpoint $I=100$~pA,
  temperature $T=13$~K).  (c) Schematic picture of alignment of the
  molecules on the Au(111) surface.
  Gold atoms in the topmost layer, second layer, and third layer are
  represented by gold, blue, and green color spheres of different
  shades.
  Black lines show
  the supercell of the $\sqrt{7} \times \sqrt{7}$ molecular
  alignment.  Upper case letters indicate the location of adsorption
  sites:  A for {\it atop} site, B for {\it bridge} site, and H for
  {\it hollow} site. Top view of an adamantane in the hollow-atop geometry is
  also illustrated (the center of the molecule is in the {\it hollow} site and
  the three bottom hydrogen atoms are in the {\it atop} site.)
  The three bottom hydrogen atoms can be seen at the bottom of Fig.~\ref{geom}(a), but
  not be seen in Fig.~\ref{geom}(c)}
\end{figure*}

Diamondoids form a class of hydrocarbon molecules composed of sp$^3$
hybridized carbon atoms. They can be regarded as small pieces of
diamond whose dangling bonds are terminated with hydrogen atoms.
Diamondoids are known to exhibit negative electron affinity
and they have possible applications as electron emitters,\cite{Drummond2005,
Yang2007} and other nanoscale devices.\cite{Wang2008}
Diamondoids have also attracted much interest because of their
possible appearance in the interstellar medium.\cite{Blake1988, Pirali2007}
The smallest diamondoid is adamantane ($\rm{C}_{10}\rm{H}_{16}$), which
has a highly symmetric cage-like structure (T$_{\rm{d}}$ point group)
illustrated in Fig.~\ref{geom}(a).  Various theoretical and
experimental results on infrared (IR) spectroscopy of both gas and
solid adamantane have
been reported in the literature.\cite{Bailey1971, Broxton1971, Wu1971,
Corn1984, Bistricic1995, Szasz1999, Jensen2004, Pirali2007,
Pirali2012} Self-assembled monolayers
of larger diamondoids than adamantane (i.~e.~tetramantane) on a
Au(111) surface system has been studied 
with scanning tunneling microscopy (STM).\cite{Wang2008, Ivan2013}
The infrared spectrum of a functionalized adamantane on a Au(111)
surface has been studied in Ref.~\onlinecite{Kitagawa2006}; however
the functionalization prevents adamantane molecules from being in
direct contact with the Au(111) surface.
Therefore a detailed characterization of the adamantane monolayer--gold
surface interaction is missing.

In this work we investigate a submonolayer of adamantane
in direct contact with a Au(111) surface. We experimentally obtained
the IR spectrum of a
self-assembled adamantane island on a Au(111) surface by using a newly
developed method which combines infrared spectroscopy and scanning
tunneling microscopy.\cite{Ivan2013} The observed
spectrum of the adamantane monolayer on Au(111)
is significantly altered with respect to the gas and
solid phase of adamantane.  
To account for this difference theoretically, 
we studied the IR spectrum within the framework of 
density functional theory (DFT) and
density functional perturbation theory (DFPT).
Our analysis reveals that intermolecular and molecule-substrate
interactions cause mixing (hybridization) of the gas phase vibrational modes.
As a result, IR-active vibrational modes of adamantane molecules on a
gold substrate are found to be considerably different than those of
the gas phase.
For example, our calculations show that the intermolecular
interaction in the adamantane monolayer
reduces the IR intensity of one of the gas-phase IR peaks by a factor of
3.5. In addition, the interaction between
adamantane molecules and the Au(111) substrate increases the IR intensity
of another gas-phase mode by a factor of 2.6 and causes a significant
redshift of $276$~cm$^{-1}$ for this mode.

This paper is organized as follows: in Sec.~\ref{sec:experiment},
we describe the details of our experiment.
In Sec.~\ref{sec:theory}
we introduce the computational methods.
In Sec.~\ref{sec:results}, we describe and analyze the theoretical and
experimental IR spectra.  In Sec.~\ref{sec:res_theory_single},
\ref{sec:res_theory_monolayer}
and \ref{sec:res_theory_gold} we study theoretically the IR spectrum of a 
single adamantane molecule, the IR spectrum of an
adamantane monolayer, and the IR spectrum of an adamantane monolayer on
a Au(111) surface, respectively. In Sec.~\ref{sec:th_exp} we compare
experimental and theoretical IR spectra.
Finally, in Sec.~\ref{sec:conclusion} we discuss our conclusions.

\section{Experimental setup}
\label{sec:experiment}

Adamantane (Sigma-Aldrich, purity $\geq$ 99\%) was deposited onto a
clean Au(111) surface from adamantane powder held 
in a vacuum chamber by exposing the gold surface to
adamantane vapor formed at room temperature. 
To achieve submonolayer molecule coverage
the adamantane vapor flux was controlled with a leak valve.  Before
the deposition, it was necessary to precool a freshly cleaned gold
crystal to 15~K to facilitate adsorption of the molecules onto
the Au(111) surface. The precooled gold crystal was then transferred
into a chamber with base pressure of $\sim 10^{-11}$~Torr where
the crystal was held in a room temperature manipulator during the
adamantane deposition which lasted for about ten minutes (pressure rose to 
$\sim 10^{-9}$~Torr when the adamantane valve was opened for deposition).
After the deposition, the sample was
immediately transferred into a homemade ultra-high vacuum variable
temperature STM operating at T = 13--15~K for STM
surface characterization.  The adamantane molecules on Au(111) were
observed to self-assemble into hexagonally packed molecular islands with
a lattice constant of 7.5$\pm$0.2~{\AA}.  Figure~1(b) shows a typical
STM image of an adamantane island on a Au(111) surface.

IR absorption spectra of adamantane submonolayers on the Au(111)
surface were obtained by using a recently developed technique referred
to as infrared scanning tunneling microscopy (IRSTM).\cite{Ivan2013} IRSTM
employs an STM tip in tunneling mode as a sensitive detector to
measure the thermal expansion of a sample due to molecular absorption
of monochromatic IR radiation. The surface  thermal expansion of the sample,
recorded as a function of IR frequency, yields the IR molecular
absorption spectrum. Frequency-tunable IR excitation of the
samples was achieved by using a homemade tunable mode-hop-free laser
source based on a singly resonant optical parametric
oscillator.\cite{Hong2012} The detailed description of the IRSTM
setup and the discussion of its performance are given
elsewhere.\cite{Ivan2013}

\section{Theoretical calculations}
\label{sec:theory}

In this section, we describe the computational methods used in this
work.

\subsection{Geometry of adamantane on Au(111)}
\label{sec:geometry}

We start with a discussion of the orientation of an adamantane monolayer on
the Au(111) surface as shown in Fig.~\ref{geom}.
Based on the STM topography shown in Fig.~\ref{geom}(b), we model the
molecular arrangement on a Au(111) surface.  In our model, the
adamantane molecules are arranged in a $\sqrt{7} \times \sqrt{7}$
structure as shown in Fig.~\ref{geom}(c).  The intermolecular distance
in this model is 7.40~{\AA}, which is close to the observed
intermolecular distance 7.5~$\pm$~0.2~{\AA}.  In our calculations we
place the adamantane molecules so that the three-fold axis of the molecule
is perpendicular to the surface, with three bottom hydrogen atoms facing down
toward the Au(111) surface.
Because of its three-fold symmetry, this configuration is compatible
with the hexagonal self-assembled island seen in the STM topography
Fig.~\ref{geom}(b). 

We determine the most stable adsorption geometry of adamantane by
computing the total energy for various adsorption sites.  
We perform \textit{ab initio} total energy calculations within the
framework of DFT\cite{Hohenberg1964, Kohn1965} to understand the
properties of an adamantane monolayer on a Au(111) surface.  We use
the local density approximation (LDA) for the exchange and correlation
energy functionals based on the quantum Monte-Carlo results of
Ceperley and Alder\cite{Ceperley1980} as parameterized by Perdew and
Zunger.\cite{Perdew1981} Vanderbilt ultrasoft
pseudopotentials\cite{Vanderbilt1990} are adopted in combination with
a plane wave basis with cut off energies of 30 and 360 Ry for the
wavefunctions and charge density, respectively. A Brillouin zone
integration is done on an 8 $\times$ 8 $\times$ 1 uniform $k$-grid.
Gaussian smearing with a 0.01~Ry width is used for the calculation of
metallic systems. We use the Quantum ESPRESSO
package\cite{Giannozzi2009} to perform DFT calculations. We also use
XCrySDen\cite{Kokalj2003} to visualize the results.

We model the Au(111) surface by a finite slab with a thickness of
seven gold layers with seven gold atoms in each layer in
a supercell geometry\cite{Cohen1975}.
The primitive unit cell of the slab includes one adamantane
molecule. Therefore, in total the primitive unit cell contains 75
atoms. The width of the vacuum region is 15.5~{\AA}.

Binding energies and molecule-surface distances of four different optimized
geometries are listed in Table~\ref{tb:energy}.
The name of each geometry is
based on the position of the center of the molecule 
and the positions of the three bottom hydrogen atoms shown in Fig.~\ref{geom}(a).
In the most stable geometry (hollow-atop) the
center of the molecule is on the hollow site and three
hydrogen atoms are close to the gold atoms in the topmost layer
(see Fig.~\ref{geom}(c)). 
The calculated
distance between the bottom hydrogen atoms and the gold surface is
2.29~{\AA} after optimization of the atomic coordinates.

\begin{table}
\caption{\label{tb:energy}
    Theoretical binding energies and surface-molecular distances of several
    adsorption geometries obtained within the LDA approximation.
    The binding energies in the Table are computed as 
    the sum of the total energy of the gold slab and the isolated adamantane molecule
    minus the total energy of an adamantane adsorbed on a Au(111) surface system.
    The convention for the binding sites of Au(111) ({\it atop},
    {\it bridge}, {\it hollow}) is as in Fig.~\ref{geom}(c).
    In our naming convention (for example {\it atop-bridge}),
    the first part ({\it atop}) represents the position of the center
    of the adamantane molecule, and second part ({\it bridge})
    represents position of the three bottom hydrogen atoms of
    adamantane.}
\begin{ruledtabular}
\begin{tabular}{ccc}
Geometry      & Distance & Binding energy \\
              &  {\AA}   & meV/molecule \\
\hline
atop-bridge   & 2.25      &  379 \\
atop-hollow   & 2.25      &  384 \\
hollow-atop   & 2.29      &  490 \\
hollow-hollow & 2.32      &  355 \\
\end{tabular}
\end{ruledtabular}
\end{table}

It is well known that LDA energy functionals do not correctly
describe long-range interactions such as the 
van der Waals interaction. Thus we
also cross-check our results with van der Waals density functionals
(vdW-DFs).\cite{Rydberg2003, Lee2010, Cooper2010, Hamada2011, Li2012} 
In particular, we use an improved version\cite{Lee2010} of the nonlocal
vdW correlation functional together with Cooper
exchange\cite{Cooper2010}.
We optimize the four different structures to compare the
energetics with those of the LDA. 
After the structural optimization with a vdW
functional, the hollow-atop geometry remains the most stable configuration although
the binding energy differences are reduced compared to the LDA 
(the binding energy difference between the hollow-atop geometry and the atop-bridge 
geometry is reduced from 111~meV/molecule (in LDA) to 44~meV/molecule).
In addition, the distance between the molecule and surface using the vdW
functional is increased from 2.29~{\AA} (in LDA) to 2.45~{\AA} in the hollow-atop geometry.
Since both of these changes are not large, we expect that the
use of the vdW-DF functionals throughout this work does not 
qualitatively affect the computed IR spectra.

\subsection{Phonon and IR intensity calculation}
\label{sec:ph_ir}

Next, we describe the methods with which we compute the frequency and
infrared intensity of the adamantane vibration modes, in
various environments (gas, monolayer, and on the Au(111) substrate).

For this purpose we use density functional perturbation theory as described in
Ref.~\onlinecite{Baroni2001}. All calculations are done only in the
Brillouin zone center point ($\Gamma$). 
We define the phonon effective charge of the $j^{\rm{th}}$ phonon branch as
\begin{equation}
  Q^{\alpha} (\omega_j) =\sum_{\beta, s}
    Z_{s}^{\alpha \beta} U_{s}^{\beta}(\omega_j).
  \label{eq:charge}
\end{equation}
Here $\mathbf{Z}$ is the Born effective charge tensor,
$\mathbf{U}(\omega_j)$ is the eigendisplacement vector, and $\omega_j$
is the phonon frequency of the $j^{\rm{th}}$ phonon branch.
Cartesian vector components are represented by $\alpha$ and $\beta$,
while $s$ represents the atom index. The Born effective charge
$Z_s^{\alpha \beta}$ is defined as the first derivative of the force
$F_s^{\beta}$ acting on an atom $s$ with respect to the electric
field $E_{\alpha}$,
\begin{equation}
Z_s^{\alpha \beta} = \frac{\partial F_s^{\beta}}{\partial E_{\alpha}}.
\label{eq:born}
\end{equation}
The IR intensity of the $j^{\rm{th}}$ phonon branch can be computed from the
phonon effective charge\cite{Porezag1996}
\begin{equation}
  I_{\rm_{IR}}(\omega_j) = \sum_{\alpha} \left| 
  Q^{\alpha} (\omega_j) \right|^2.
\label{eq:IR}
\end{equation}
Once we obtain the IR intensities $I_{\rm_{IR}}(\omega_j)$, we model
the IR spectrum at any frequency $\omega$ by assuming a Lorentzian lineshape
with a constant linewidth of 10~$\rm{cm}^{-1}$ (full width at half
maximum).
In general, all nine Cartesian components of the Born effective charge
must be calculated to obtain the IR intensities.  However, 
on the metallic surface, one can focus only on the components of the
electric field  perpendicular to the 
surface ($\alpha=z$).\cite{Greenler1966, Hexter1979,Sheppard1984} 

To compute the Born effective charge, we use a finite-difference
approximation of Eq.~\ref{eq:born}, and we apply the electric field using
a saw-tooth like potential in the direction perpendicular to the slab
($z$). Therefore we can obtain the Born effective charge by dividing the
force induced by the electric field ($\Delta F$) with the strength of
the electric field ($E_z$). 

Using this method we calculate the IR spectra of a single molecule,
of a molecular monolayer without a substrate, and of molecules on a Au(111)
substrate.
To track the changes in the IR spectrum due to intermolecular
interaction and due to substrate effects, we perform the following
interpolation procedure. We define an interpolated dynamical matrix
$\mathbf{D}_{\rm{int}}$ and Born effective charge
$\mathbf{Z}_{\rm{int}}$ between any two configurations A and B as
\begin{align}
\mathbf{D}_{\rm{int}} &= (1-\lambda) \mathbf{D}_{\rm{A}} + \lambda
\mathbf{D}_{\rm{B}}, \notag \\
\mathbf{Z}_{\rm{int}} &= (1-\lambda) \mathbf{Z}_{\rm{A}} + \lambda
\mathbf{Z}_{\rm{B}}.
\label{eq:interpolation}
\end{align}
Here $\mathbf{D}_{\rm{int}}=\mathbf{D}_{\rm{A}}$ for $\lambda=0$ and
$\mathbf{D}_{\rm{int}}=\mathbf{D}_{\rm{B}}$ for $\lambda=1$ and is
continuously tuned from A to B for $0 < \lambda < 1$ (same for
$\mathbf{Z}_{\rm{int}}$). 
Diagonalizing $\mathbf{D}_{\rm{int}}$ for each $\lambda$ and using
interpolated $\mathbf{Z}_{\rm{int}}$ we obtain interpolated IR
intensity for each $\lambda$ with $0 < \lambda < 1$.

For the start and end configurations A and B, we use 
either no subscript, subscript M, or subscript Au
to denote either an isolated molecule, molecules in an isolated molecular monolayer,
or in a monolayer on a Au(111) substrate, respectively. We first
interpolate the spectra from the isolated molecule phase to the
monolayer phase (M) to determine the effect of the intermolecular
interactions. Finally to estimate the effect of the
molecule-substrate interactions we interpolate the spectrum of the
monolayer phase (M) to the case of molecules on the gold substrate (Au).

In the interpolation procedure, we take into account only dynamical
matrix elements of carbon and hydrogen atoms (neglecting the
displacements of gold atoms). We find that the effect of gold atom
displacements on vibrational frequencies is
only 0.3~$\rm{cm}^{-1}$ and on the IR intensity less than
10\%.

In addition to the interpolation method we also quantitatively
analyze the similarity of eigendisplacement vectors between various
configurations.
We do this by computing the norm of the inner product $\left|
\braket{u_{i}^{\rm{A}}|u_{j}^{\rm{B}}}\right|^2$ between $i^{\rm{th}}$
phonon eigendisplacement vector $\bra{u_{i}^{\rm{A}}}$ in
configuration A, and $j^{\rm{th}}$ phonon eigendisplacement vector
$\ket{u_{j}^{\rm{B}}}$ in configuration B.
We use the same subscript convention as for the interpolation
procedure (Au, M, or no subscript for molecules on the
substrate, the molecular monolayer, or single molecule case
respectively). 
Since an adamantane molecule consists of 78
phonon modes, we simplify the analysis by only considering inner
products between the 16 predominantly C--H stretching modes.

\section{Results}
\label{sec:results}

This section is organized as follows. 
We present our
experimentally obtained IR spectra of an adamantane submonolayer
on a Au(111) surface in
Sec.~\ref{sec:res_experiment}. Next, in Sec.~\ref{sec:res_theory} we
analyze theoretically obtained IR spectra. Finally, in
Sec.~\ref{sec:th_exp} we compare theory and experiment.

\subsection{Experimental IR spectra}
\label{sec:res_experiment}

\begin{figure}[h!]
\includegraphics[width = 8.5cm]{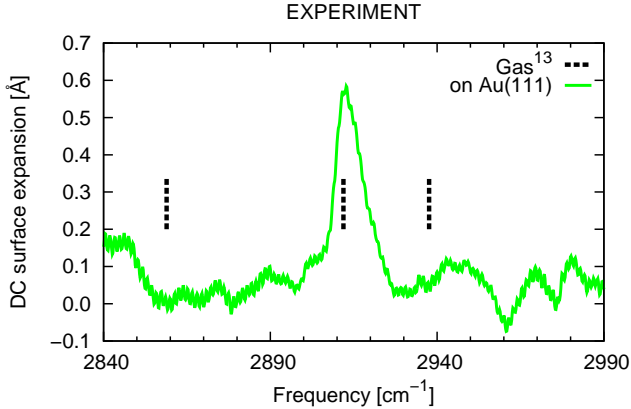}
\caption{\label{STMIR} (Color online) The green line shows the
  experimentally observed spectrum (averaged over 15 sweeps) of 0.8 ML
  of adamantane on a Au(111) surface with gold baseline signal
  subtracted. The IR absorption peaks are seen at 2846 $\pm$ 2 and
  2912 $\pm$ 1 $\rm{cm}^{-1}$. The vertical black dashed lines show the
  IR peak positions of an adamantane molecule in the gas phase as listed in Table
  \ref{tb:gas} (2859, 2912, 2938 $\rm{cm^{-1}}$).}
\end{figure}

Figure~\ref{STMIR} shows an experimentally measured IRSTM spectrum
(green line) of
0.8~ML of adamantane adsorbed on a Au(111) surface.
The spectrum was obtained by
measuring the STM Z-signal under constant-current feedback conditions
while sweeping the IR excitation from 2840~$\rm{cm}^{-1}$ to
2990~$\rm{cm}^{-1}$ 
(the spectrum shown was averaged over 15 frequency
sweeps and background-corrected by subtracting a linear fit
to the estimated bare gold contribution to the spectrum).  Two IR
absorption peaks for adamantane/Au(111) can clearly be seen at
2846~$\pm$~2 and 2912~$\pm$~1~$\rm{cm}^{-1}$ in Fig.~\ref{STMIR}.  The
other small peaks seen in Fig.~\ref{STMIR} are not reproducible
and thus we are not able to unambiguously relate them to the
adamantane absorption.  The black dashed lines in Fig.~\ref{STMIR} 
show the IR peak
positions of an adamantane molecule in the gas phase.\cite{Pirali2012} Comparing the green curve and
the black dashed line in Fig.~\ref{STMIR}, it is clear that the IR spectrum of
adamantane on the Au(111) surface is considerably different 
from the gas phase spectrum.

\subsection{Theoretical analysis of IR spectra}
\label{sec:res_theory}

To understand the origin of IR spectrum modification of adamantane on
Au(111) we compute IR spectra for an isolated adamantane molecule, for 
an adamantane monolayer, and for an adamantane monolayer placed on 
the Au(111) substrate. We discuss these
three cases in the following three subsections of the paper.

\subsubsection{Single molecule}
\label{sec:res_theory_single}

\begin{table}
  \caption{ \label{tb:gas} Calculated vibrational frequencies (in
    cm$^{-1}$) and phonon effective charge $Q^{z}$ 
    (in elementary charge $e$) of 
    an isolated adamantane molecule from
    2850~cm$^{-1}$ to 2950~cm$^{-1}$. Experimental
    values are shown for the gas phase\cite{Pirali2012} and solution phase.\cite{Bistricic1995}
    We label phonon modes by $\omega_1$--$\omega_7$ with one label
    corresponding to one irreducible representation (irrep). 
    Only $\rm{T}_2$ modes can be observed in the gas IR
    spectroscopy. $\rm{A}_1$, E, and $\rm{T}_2$ are Raman active,
    while $\rm{T}_1$ modes are inactive (both in IR and Raman).
  }
\begin{ruledtabular}
\begin{tabular}{ccccccc}
	&  \multirow{2}{*}{Irrep} &\multirow{2}{*}{Frequency} & \multirow{2}{*}{$Q^{z}$} & \multicolumn{2}{c}{Prev.~experiment}\\
\cline{5-6}
           &              &      &  & Ref.~\onlinecite{Pirali2012}\footnotemark[1] & Ref.~\onlinecite{Bistricic1995}\footnotemark[2]\\
\hline
$\omega_1$ & $\rm{A}_1$   & 2924      &      -      & -    & 2913 \\
$\omega_2$ & $\rm{A}_1$   & 2891      &      -      & -    & 2857 \\
$\omega_3$ & $\rm{E}$     & 2892      &      -      & -    & 2900 \\
$\omega_4$ & $\rm{T}_1$   & 2938      &      -      & -    & -    \\
$\omega_5$ & $\rm{T}_2$   & 2940      &    0.25     & 2938 & 2950 \\
$\omega_6$ & $\rm{T}_2$   & 2918      &    0.32     & 2912 & 2904 \\
$\omega_7$ & $\rm{T}_2$   & 2892      &    0.19     & 2859 & 2849 \\
\end{tabular}
\footnotetext[1]{IR spectroscopy of gas phase adamantane.}
\footnotetext[2]{IR and Raman spectroscopy of adamantane solution.
  Mode assignment is based on Ref.~\onlinecite{Szasz1999}.}
\end{ruledtabular}
\end{table}

We compute the vibrational frequencies and IR intensities of an isolated
single molecule of adamantane by placing the molecule in a large unit
cell (length of each side is 16~{\AA})
to minimize the interaction between periodic replicas. 
The calculated vibrational frequencies
of the C--H bond stretching modes (and phonon effective charge) 
are listed in Table~\ref{tb:gas} and compared with the
experimental frequencies from the literature.\cite{Pirali2012,Bistricic1995}
Our calculation reproduces quite well the vibrational frequencies as
compared to the experimental data. We find the largest discrepancy of
$\sim30$~cm$^{-1}$ for two modes (labeled $\omega_2$ and
$\omega_7$ in Table~\ref{tb:gas}), while the discrepancy for the other modes
is only $\sim10$~cm$^{-1}$.

\begin{figure}[h!]
\includegraphics[width = 8.5cm]{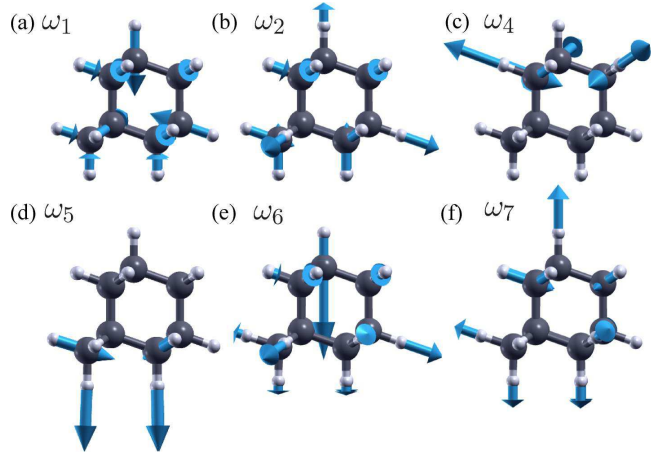}
\caption{\label{gdisp} (Color online) Eigendisplacement vectors of (a)
  $\omega_1$, (b) $\omega_2$, (c) $\omega_4$, (d) $\omega_5$, (e)
  $\omega_6$, and (f) $\omega_7$ modes of an isolated adamantane
  molecule (notation is from Table~\ref{tb:gas}).
  Gray and white spheres represent carbon and hydrogen
  atoms, respectively. Blue arrows indicate the displacements of the atoms
  (mostly hydrogen atoms).  
  Some of the atoms and eigendisplacements are overlapping since the
  molecule is shown from a high-symmetry direction.}
\end{figure}

There are 16 C--H stretching modes in adamantane (equal to the number of C--H bonds).
Out of 16 modes, there are three modes corresponding to the $\rm{T}_2$
irreducible representation, one $\rm{T}_1$, one $\rm{E}$, and two
$\rm{A}_1$ modes.  In the highly symmetric adamantane molecule (point
group T$\rm{_d}$) only these three $\rm{T}_2$ modes are IR active while
other C--H stretching modes are IR inactive.
The IR active modes
have the following approximate characters: asymmetric
stretching of the C--H$_2$ bonds (labeled $\omega_5$ in 
Table~\ref{tb:gas}), symmetric C--H stretch mode ($\omega_6$), and
symmetric C--H$_2$ stretch mode ($\omega_7$). Their
eigendisplacement vectors are also shown in
Figs.~\ref{gdisp}(d), (e), and (f).

\begin{figure}[h!]
\includegraphics[width = 8.5cm]{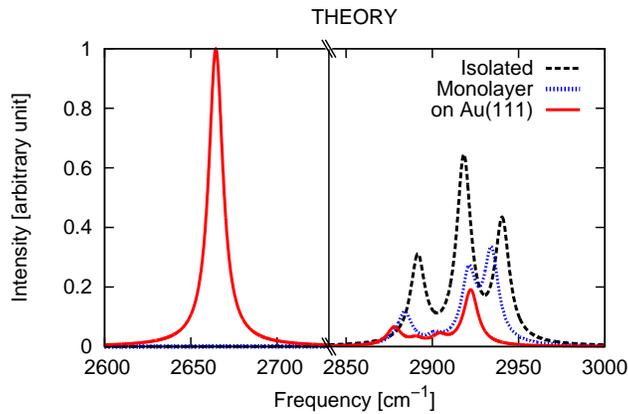}
\caption{\label{IR} (Color online) Calculated IR spectrum of an
  isolated adamantane molecule (black dashed line), an adamantane monolayer
  (blue dotted line), and an adamantane monolayer on
  Au(111) (red solid line).  
  The IR spectra are displayed in arbitrary units (chosen so that the intensity of 
  the IR peak around 2650~cm$^{-1}$ equals 1.0).
  The black vertical line in the figure separates lower and higher
  frequency regimes (there are no IR active modes in the intermediate
  regime between 2730 and 2840~cm$^{-1}$).
  Here we show only the IR spectra close to the C--H stretching mode
  frequencies (the closest IR mode not corresponding to C--H stretching
  is below 1500~cm$^{-1}$).}
\end{figure}

The calculated IR spectrum of an isolated adamantane molecule is indicated in
Fig.~\ref{IR} by a black dashed line.
Here we compute only the $\alpha = z$ component of the IR intensities (Eq.~\ref{eq:IR})
to simplify the comparison with the IR intensities of 
adamantane in molecular monolayer and monolayer on the Au(111) surface.
We obtain three substantial IR absorption peaks of a single molecule
adamantane in the frequency region from 2850 to 2950~cm$^{-1}$. 
The highest, middle, and lowest frequency peaks correspond to 
the $\omega_5$, $\omega_6$, and $\omega_7$ modes, respectively.
The $\omega_6$ peak has the largest IR intensity, followed by the
$\omega_5$ peak.
This order of the IR intensities is consistent with a previous DFT result\cite{Jensen2004}.

\subsubsection{Adamantane monolayer}
\label{sec:res_theory_monolayer}

\begin{table}[h!]
  \caption{ \label{tb:mono} Calculated vibrational frequencies (in
    cm$^{-1}$) and phonon effective charges $Q^{z}$ 
    (in elementary charge $e$) of an adamantane monolayer.
    We label monolayer phonon modes by $\omega_1^{\rm M}$--$\omega_6^{\rm M}$ with one label
    corresponding to one irreducible representation (irrep). 
    Norm of the overlap between gas phase phonon eigenvector ($\bra{u_j}$) and
    monolayer phonon eigenvector ($\ket{u_i^{\rm M}}$) is also
    shown, indicating nature of the hybridization of the gas phase phonons into the
    monolayer phase phonons. In the table we neglect all overlaps
    whose norm is smaller than 0.1.} 
\begin{ruledtabular}
\begin{tabular}{cccccccccccc}
& \multicolumn{7}{c}{Gas phase overlap} & \quad & \multirow{2}{*}{Irrep} & \multirow{2}{*}{Frequency} & \multirow{2}{*}{$Q^z$}  \\
\cline{2-8}
& $\omega_1$ & $\omega_2$ & $\omega_3$ & $\omega_4$ & $\omega_5$ & $\omega_6$ & $\omega_7$    &&           &           &          \\
\hline
$\omega_1^{\rm{M}}$ & 0.36       & .          & .          & .          & .          & 0.63       &  .    &&
$\rm{A}_1$ & 2921 &  0.19  \\
$\omega_2^{\rm{M}}$ & .          & 0.95       & .          & .          & .          & .          &  .    &&   
$\rm{A}_1$ & 2879 &  0.03  \\
$\omega_3^{\rm{M}}$ & .          & .          & .          & 0.99       & .          & .          &  .    &&  
$\rm{A}_2$ & 2931 &     -   \\
$\omega_4^{\rm{M}}$ & .          & .          & .          & .          & 0.90       & .          &  .    &&  
$\rm{A}_1$ & 2934 &  0.22  \\
$\omega_5^{\rm{M}}$ & 0.58       & .          & .          & .          & .          & 0.33       &  .    &&
$\rm{A}_1$ & 2902 &  0.06  \\
$\omega_6^{\rm{M}}$ & .          & .          & .          & .          & .          & .          &  0.94 &&
$\rm{A}_1$ & 2884 &  0.13  \\
\end{tabular}
\end{ruledtabular}
\end{table}

Before analyzing the IR spectrum of an adamantane monolayer on a Au(111)
surface, we first analyze the IR spectrum of a free adamantane
monolayer (with the same intermolecular distance as in the monolayer
on the gold substrate).

Placing adamantane in the monolayer arrangement (see Sec.~\ref{sec:geometry}) lowers the symmetry of
the system from $\rm{T_d}$ (in the gas phase) to $\rm{C_{3v}}$.
This symmetry reduction is followed by splitting of the threefold
degenerate $\rm{T}_2$ representation (in $\rm{T}_{\rm{d}}$) to a twofold
$\rm{E}$ representation and a onefold $\rm{A}_1$ representation.
The basis functions of the E representation are $x$ and $y$, therefore
they have no IR activity in the $z$ direction (perpendicular to the
monolayer).
In contrast, the modes with A$_1$ representation are active along the
$z$ direction.
In addition, symmetry reduction to the adamantane monolayer splits the 
$\rm{T}_1$ representation into $\rm{E}$ and $\rm{A}_2$, both of which
are IR inactive in the $z$ direction ($\rm{E}$ is active in $x$ and
$y$).
Therefore, the adamantane monolayer has in total five C--H stretching modes
that are IR active along the $z$ direction.
We label these five modes corresponding to the A$_1$
representation with $\omega_1^{\rm{M}}$, $\omega_2^{\rm{M}}$, 
$\omega_4^{\rm{M}}$, $\omega_5^{\rm{M}}$, and $\omega_6^{\rm{M}}$.

\begin{figure}[h!]
\includegraphics[width = 8.5cm]{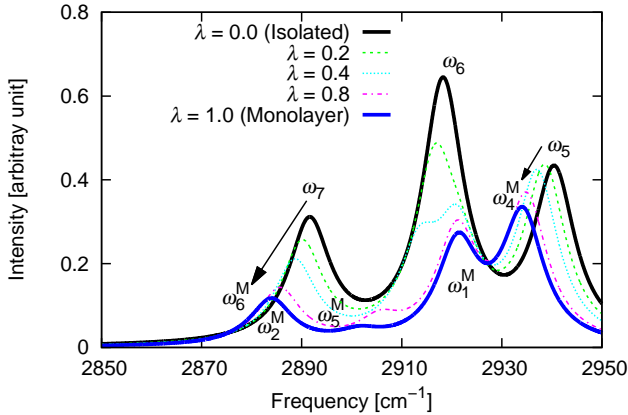}%
\caption{\label{gtm} (Color online) Interpolated IR spectra between
  the single molecule adamantane (solid black line) and the adamantane
  monolayer (solid blue line). The dashed ($\lambda=0.2$), dotted
  ($\lambda=0.4$), and chain ($\lambda=0.8$) lines are interpolated
  spectra between two cases (using techniques from
  Sec.~\ref{sec:ph_ir}). The arrows indicate the peaks evolution from the
  isolated molecule case to the monolayer case.}
\end{figure}

The blue dashed line in Fig.~\ref{IR} shows the calculated
IR spectrum of the adamantane monolayer. 
The changes in the IR spectrum of the adamantane
monolayer compared to the single molecule are presented in more detail
in Fig.~\ref{gtm} and in Table~\ref{tb:mono}.
Figure~\ref{gtm} shows interpolated IR spectra between a single
molecule and a molecular monolayer case (using the interpolation method
described in Sec.~\ref{sec:ph_ir}). Table~\ref{tb:mono} shows the
calculated vibrational frequencies, phonon effective 
charges, and inner products
between phonon eigenvectors of gas and monolayer adamantane.

\begin{figure}[h!]
\includegraphics[width = 8.5cm]{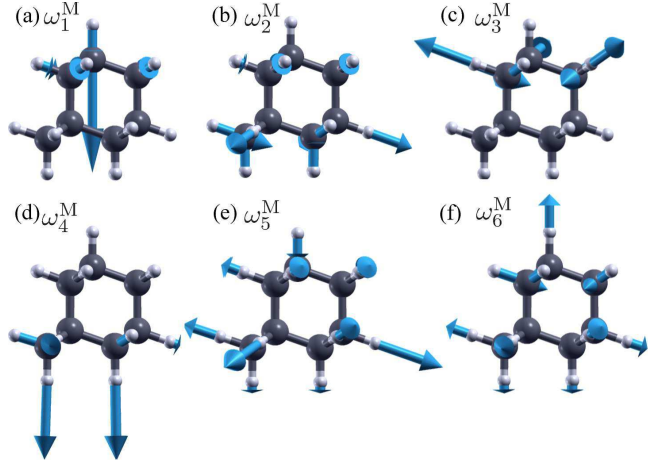}
\caption{\label{mdisp} (Color online) Phonon eigendisplacement vectors of (a)
  $\omega_1^{\rm{M}}$, (b) $\omega_2^{\rm{M}}$, (c)
  $\omega_3^{\rm{M}}$, (d) $\omega_4^{\rm{M}}$, (e)
  $\omega_5^{\rm{M}}$, and (f) $\omega_6^{\rm{M}}$ modes of an
  adamantane monolayer without a gold substrate.}
\end{figure}

From Fig.~\ref{gtm} and the inner products in Table~\ref{tb:mono}, we find
that there is one-to-one correspondence between the $\omega_2$,
$\omega_4$, $\omega_5$, $\omega_7$ modes of a single molecule
adamantane phase and the $\omega_2^{\rm{M}}$, $\omega_3^{\rm{M}}$,
$\omega_4^{\rm{M}}$, and $\omega_6^{\rm{M}}$ modes of the monolayer
adamantane phase, respectively. This correspondence is also evident in the
similarity of the phonon eigendisplacement vectors of these two phases
(compare Figs.~\ref{gdisp}(b), (c), (d), (f) and
Figs.~\ref{mdisp}(b), (c), (d), (f)).  
Three of these modes ($\omega_2$, $\omega_5$, and $\omega_7$) are
redshifted by about 10~cm$^{-1}$ in the monolayer phase (see arrows in
Fig.~\ref{gtm})
which is close to the redshifts found in the solid phase
diamondoids.\cite{Pirali2007}
The IR inactive mode $\omega_{4}$ is
redshifted by 7~cm$^{-1}$ in the monolayer phase.

The remaining IR active C--H stretching modes in the adamantane monolayer
($\omega_1^{\rm{M}}$ and $\omega_5^{\rm{M}}$) result from a strong
mixing of the $\omega_1$ and $\omega_6$ modes in the gas phase (the inner
products shown in Table~\ref{tb:mono} are between 0.33 and 0.63)
which also affects their eigendisplacement patterns
(compare Figs.~\ref{gdisp}(a) and (e) and Figs.~\ref{mdisp}(a) and (e)).
In addition, the $\omega_5^{\rm{M}}$ mode is redshifted by about
20~cm$^{-1}$ with respect to the $\omega_1$ and $\omega_6$ mode in the
gas phase.

\subsubsection{Monolayer on Au(111)}
\label{sec:res_theory_gold}

Finally, we study the vibrational properties of the adamantane
monolayer on a Au(111) surface. Introduction of the Au(111) surface further
reduces the symmetry of the system from $\rm{C_{3v}}$ (in the monolayer) 
to $\rm{C_3}$.
Due to symmetry reductions, both the IR active A$_1$ mode and the IR inactive
A$_2$ mode having $\rm{C_{3v}}$ symmetry (monolayer) are changed to the IR
active A representation having C$_3$ symmetry. The IR inactive E modes,
remains inactive in the C$_3$ symmetry along the $z$ direction.
The red solid line in Fig.~\ref{IR} shows the
calculated IR spectrum of a molecular monolayer on a gold substrate.
Vibrational frequencies and phonon effective charges are shown in
Table~\ref{tb:gold}. The most notable difference compared to the spectrum
of the isolated adamantane monolayer and gas phase molecule is the significant redshift of
one of the modes $\omega_6^{\rm{Au}}$ (to 2664~cm$^{-1}$) with a sizable increase in the 
phonon effective charge (0.40~$e$).

\begin{table}[h!]
  \caption{\label{tb:gold} Vibrational frequencies (in cm$^{-1}$), 
    phonon effective charges $Q^{z}$ (in elementary charge $e$), and inner products
    of C--H stretching modes of adamantane molecules on a Au(111)
    surface.
    Norm of the overlap between adamantane monolayer 
    phase phonon eigenvector ($\bra{u_j^{\rm M}}$) and
    monolayer on gold phonon eigenvector ($\ket{u_i^{\rm Au}}$) is also
    shown, indicating the nature of the hybridization of the gas phase phonons into the
    monolayer phase phonons. In the Table we neglect all overlaps
    whose norm is smaller than 0.1.}
\begin{ruledtabular}
\begin{tabular}{ccccccccccc}
  & \multicolumn{6}{c}{Monolayer phase overlap} &\quad&
  \multirow{2}{*}{Irrep} & \multirow{2}{*}{Frequency} &
  \multirow{2}{*}{$Q^z$} \\
  \cline{2-7}
  & $\omega_1^{\rm M}$ & $\omega_2^{\rm M}$ & $\omega_3^{\rm M}$ &
  $\omega_4^{\rm M}$ & $\omega_5^{\rm M}$ & $\omega_6^{\rm M}$ && & & \\
  \hline
  $\omega_1^{\rm{Au}}$ & 1.00 &   .  &  .   &  .   &   .  &  .   && A &  2922 &  0.17 \\
  $\omega_2^{\rm{Au}}$ &   .  & 0.35 &  .   &  .   & 0.27 & 0.33 && A &  2878 &  0.10 \\ 
  $\omega_3^{\rm{Au}}$ &   .  &   .  & 1.00 &  .   &   .  &  .   && A &  2935 &  0.01 \\
  $\omega_4^{\rm{Au}}$ &   .  & 0.34 &  .   & 0.21 & 0.42 &  .   && A &  2904 &  0.06 \\
  $\omega_5^{\rm{Au}}$ &   .  &   .  &  .   &  .   & 0.29 & 0.63 && A &  2891 &  0.07 \\
  $\omega_6^{\rm{Au}}$ &   .  & 0.27 &  .   & 0.70 &   .  &  .   && A &  2664 &  0.40 \\
\end{tabular}
\end{ruledtabular}
\end{table}

\begin{figure}[h!]
 \includegraphics[width = 8.5cm]{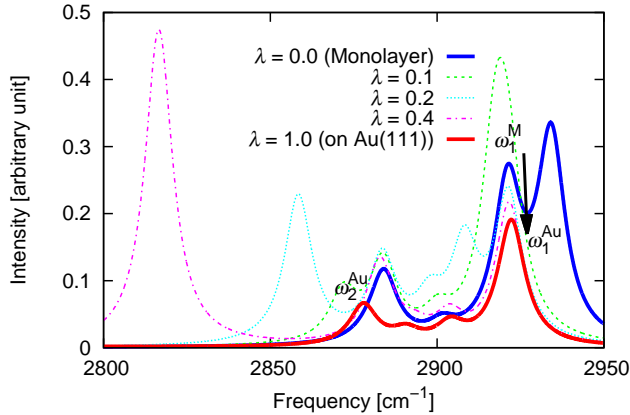}
 \caption{\label{mts} (Color online) Interpolated IR spectra going
   from an isolated molecular monolayer to a monolayer on Au(111). The
   blue solid line shows the IR spectrum of the isolated molecular monolayer,
   while the red solid line shows the IR spectrum of the monolayer on
   the Au(111) substrate. The dashed ($\lambda=0.1$), dotted
   ($\lambda=0.2$), and chain ($\lambda=0.4$) lines show
   an interpolated spectra between the two cases. The arrow shows the
   transition from $\omega_1^{\rm{M}}$ to $\omega_1^{\rm{Au}}$. We
   also indicate the peak position of the relatively weak
   $\omega_2^{\rm{Au}}$ mode. The severely redshifted peak $\omega_6^{\rm{Au}}$
   is not shown in this figure (see left panel of Fig.~\ref{IR}.)}
\end{figure}

\begin{figure}[h!]
\includegraphics[width = 8.5cm]{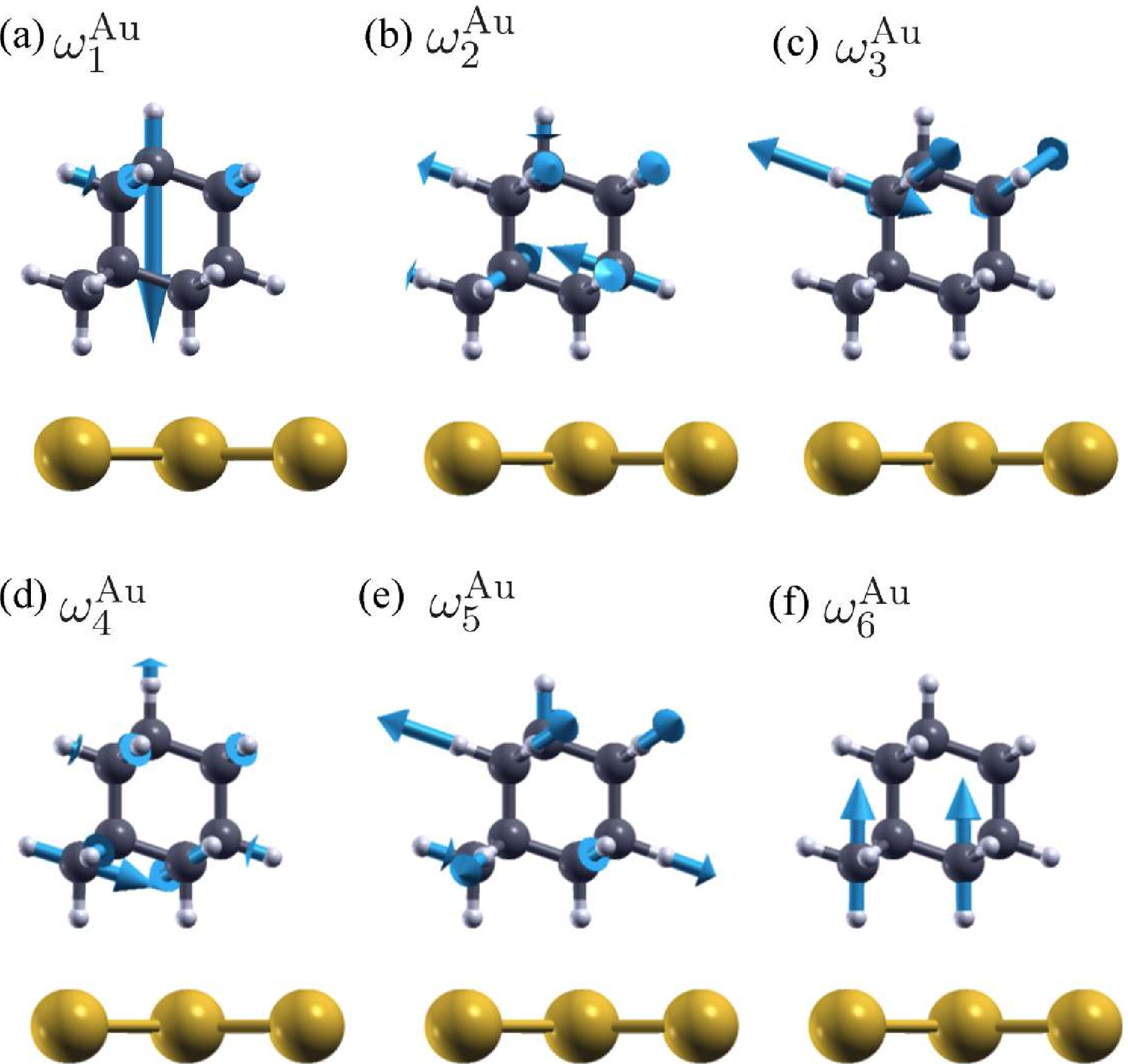}%
\caption{\label{disp} (Color online) Eigendisplacement vectors of (a)
  $\omega_1^{\rm{Au}}$, (b) $\omega_2^{\rm{Au}}$, (c)
  $\omega_3^{\rm{Au}}$, (d) $\omega_4^{\rm{Au}}$, (e)
  $\omega_5^{\rm{Au}}$, and (f) $\omega_6^{\rm{Au}}$ modes of an
  adamantane monolayer on the Au(111) surface. Yellow spheres
  illustrate Au atoms in the topmost layer of the Au(111) surface.}
\end{figure}

Figure~\ref{mts} shows the interpolated IR spectra between the
isolated monolayer case and the monolayer on Au(111).
This analysis shows that the $\omega_1^{\rm{Au}}$ mode is
nearly unaffected by the Au(111) substrate as it originates from
the stretching of a topmost C--H bond, relatively far from the Au(111)
surface. In fact, the eigendisplacement
vectors of the $\omega_1^{\rm{Au}}$ mode and $\omega_1^{\rm{M}}$ mode
are nearly the same (compare Fig.~\ref{mdisp}(a) and Fig.~\ref{disp}(a)).

Interpolation analysis of the remaining IR peaks is quite involved in
this case.
Therefore we turn to the analysis of the inner products of the phonon
modes in the isolated monolayer and the monolayer of 
on Au(111). These inner products are shown in
Table~\ref{tb:gold}. From this table we again confirm that
the $\omega_1^{\rm{Au}}$ mode originates from the $\omega_1^{\rm{M}}$
mode. In addition, we find that the $\omega_3^{\rm{Au}}$ mode has
one-to-one correspondence with the $\omega_3^{\rm{M}}$ mode.

For the remaining IR active modes ($\omega_2^{\rm{Au}}$,
$\omega_4^{\rm{Au}}$, $\omega_5^{\rm{Au}}$, and $\omega_6^{\rm{Au}}$)
we find strong influence by the monolayer-substrate
interaction. 
Analyzing Table~\ref{tb:gold} reveals that the $\omega_2^{\rm{Au}}$
mode of adamantane on the Au(111) substrate is a mixture of the
$\omega_2^{\rm{M}}$, $\omega_5^{\rm{M}}$, and $\omega_6^{\rm{M}}$
modes of the isolated monolayer phase. Similarly, the $\omega_4^{\rm{Au}}$
mode is composed of the $\omega_2^{\rm{M}}$, $\omega_4^{\rm{M}}$, and
$\omega_5^{\rm{M}}$ modes, while the $\omega_5^{\rm{Au}}$ mode comes mostly
from the $\omega_6^{\rm{M}}$ mode with some admixture of the 
$\omega_5^{\rm{M}}$ mode.  
The IR intensity of modes $\omega_2^{\rm{Au}}$, $\omega_4^{\rm{Au}}$
and $\omega_5^{\rm{Au}}$  are relatively small since these
modes include only a small amount of C--H bond stretching
perpendicular to the surface (see phonon eigendisplacements in
Fig.~\ref{disp}).

On the other hand, Table~\ref{tb:gold} shows a rather remarkable change
in the frequency and IR intensity of the $\omega_6^{\rm{Au}}$ mode.
This mode is a mixture of the $\omega_4^{\rm{M}}$ (2934~cm$^{-1}$) and
$\omega_2^{\rm{M}}$ (2879~cm$^{-1}$) modes of the isolated monolayer phase and
its frequency is red shifted to 2664~cm$^{-1}$. Furthermore, its IR
intensity is increased from 0.05~$e^2$ to 0.16~$e^2$ (effective charge 
in Table~\ref{tb:gold} changes from 0.22~$e$ to 0.40~$e$) compared to the
$\omega_4^{\rm{M}}$ mode.
The $\omega_6^{\rm{Au}}$ mode consists of the
in-phase perpendicular vibration of the three bottom hydrogen atoms near the
Au(111) surface (see Fig.~\ref{disp}(f)),
therefore it is not unexpected that this mode will be significantly
affected by the Au(111) surface.
Comparison of the charge density of the isolated adamantane monolayer to the charge
density of the monolayer on the Au(111) surface reveals that
the molecule-surface interaction reduces the electron charge density on
the adamantane C--H bonds.
We speculate that this reduction of charge density within the C--H bonds is
responsible for the decrease in the $\omega_6^{\rm{Au}}$ mode frequency as
well as increase of its effective charge.

\subsection{Comparison of theory and experiment} 
\label{sec:th_exp}

\begin{figure}\includegraphics[width = 8.5cm]{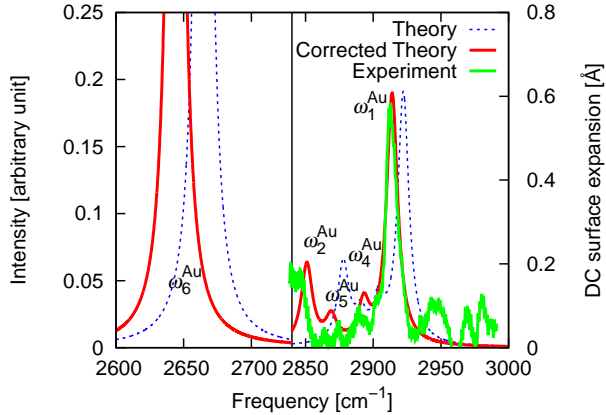}%
  \caption{\label{Cor} (Color online) Uncorrected (blue dashed line)
    and corrected (red solid line) theoretical IR spectra and
    experimentally observed IR spectrum (green solid line). The
    frequency region between 2730~cm$^{-1}$
    and 2840~cm$^{-1}$ is not shown.
    The vertical scale is chosen so that 
    the theoretical and experimental
    peaks around 2850~cm$^{-1}$ have almost the same height.
    Left and right vertical axes correspond to theoretical 
    and experimental IR intensity.  Corrected theoretical values of
    $\omega_1^{\rm{Au}}$, $\omega_2^{\rm{Au}}$, $\omega_4^{\rm{Au}}$,
    $\omega_5^{\rm{Au}}$, and $\omega_6^{\rm{Au}}$ modes are 2914,
    2851, 2893, 2869, and 2644~cm$^{-1}$, respectively.}
\end{figure}

Figure~\ref{Cor} shows a comparison of the experimental (green line)
and theoretical (dashed blue line) IR spectra for an adamantane
monolayer on the Au(111) surface. We find good qualitative agreement between
the two spectra, both in the peak position and in their relative intensities
(the experimental vertical scale is chosen so that the peak height at
2846~cm$^{-1}$ matches the theoretical peak height at 2851~cm$^{-1}$).
Agreement is even better after applying a correction to the calculated
phonon frequencies (red line in Fig.~\ref{Cor}) as we describe below in
Sec.~\ref{sec:correction}.

We assign the relatively large experimentally obtained IR peak at
2912~cm$^{-1}$ to the theoretically obtained $\omega_{1}^{\rm{Au}}$
mode (2922~cm$^{-1}$, corrected frequency 2914~cm$^{-1}$, 
phonon effective charge 0.17~$e$). Furthermore, we assign 
the relatively weaker mode at 2846~cm$^{-1}$
to the theoretically obtained $\omega_{2}^{\rm{Au}}$ mode
(2878~cm$^{-1}$, corrected frequency 2851~cm$^{-1}$, 
phonon effective charge 0.10~$e$).
Remaining features in the experimental data (green line in
Fig.~\ref{Cor}) are not reproducible and therefore cannot be reliably
assigned to the additional IR phonon modes. This is consistent with
our theory, as the remaining IR active modes $\omega_{3}^{\rm{Au}}$,
$\omega_{4}^{\rm{Au}}$, and $\omega_{5}^{\rm{Au}}$ have a much
smaller phonon effective charge (from 0.01 to 0.07~$e$).

Finally, our calculation predicts the existence of a significantly
redshifted IR active mode $\omega_6^{\rm{Au}}$ at 2664~cm$^{-1}$
(corrected value is 2644~cm$^{-1}$) with 
a large phonon effective charge (0.40~$e$).
Although the frequency of this mode is currently outside 
of our experimentally attainable
frequency range (from 2840 to 2990~cm$^{-1}$),
we expect that it will be accessible to future experimental probing.

\subsubsection{Correction of the dynamical matrix}
\label{sec:correction}

Here we present the method we use to correct the DFT-LDA IR spectrum of the
adamantane monolayer on the Au(111) substrate (red line in
Fig.~\ref{Cor}). First we obtain the correction $D^{\rm corr}$ to the
calculated dynamical matrix of the adamantane {\it gas} phase so that it exactly
reproduces the experimentally
measured frequencies of adamantane gas and solution phase,
\begin{equation}
  D^{\rm corr} = \sum_i (\Delta_i^2 - 2 \omega_i \Delta_i)\ket{u_i}\bra{u_i}.
  \label{eq:correction}
\end{equation}
Here $\omega_i$ and $\ket{u_i}$ are the phonon frequencies and eigenvectors
of the original dynamical matrix, while $\Delta_i$ is the difference
between the computed and the measured adamantane gas and solution
phase frequency.
In the second step, we add this same correction matrix $D^{\rm
  corr}$ to the dynamical matrix of the adamantane monolayer on the Au(111)
surface. Finally, we use the eigenvalues and eigenvectors of the
corrected dynamical matrix to compute the corrected IR spectrum.

Our correction procedures improve the agreement of 
the calculated IR spectrum of adamantane on the Au(111) surface with
the experimental spectrum (see Fig.~\ref{Cor}).
The theoretical peak position of 
the $\omega_1^{\rm Au}$ mode is redshifted by 8~cm$^{-1}$
(from 2922~cm$^{-1}$ to 2914~cm$^{-1}$) and sits
closer to the experimental peak position (2913~$\pm$~1~cm$^{-1}$).
Similarly, the $\omega_2^{\rm Au}$ mode is redshifted by 
27~cm$^{-1}$ (from 2878~cm$^{-1}$ to 2851~cm$^{-1}$), again
closer to the experimental value (2846~$\pm$~2~cm$^{-1}$).

\section{Summary and Conclusions}
\label{sec:conclusion}

\begin{figure}\includegraphics[width = 8.5cm]{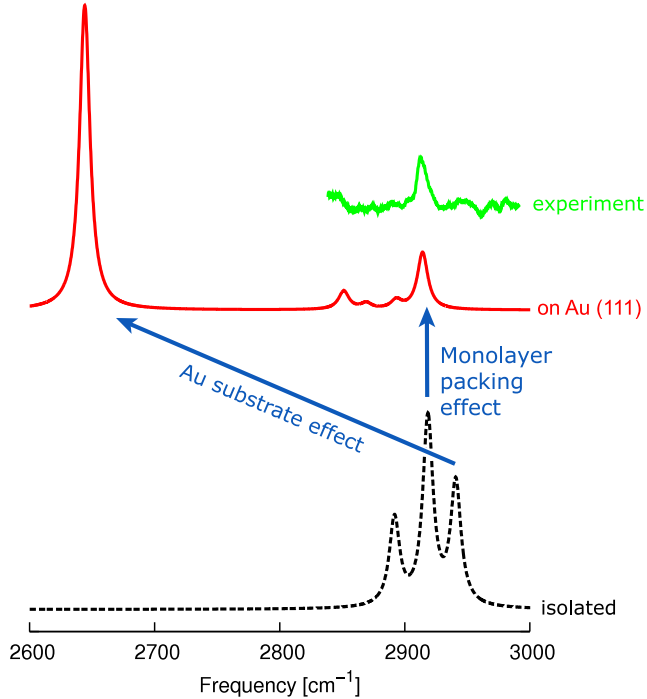}%
  \caption{\label{fig:summ} (Color online) Summary of our main
    results. There are three IR active C--H stretching modes in the
    isolated (gas phase) adamantane molecule (black dashed line). The
    interaction between the neighboring adamantane molecules in the
    monolayer (packing effect) reduces the IR intensity of the
    central IR active mode by a factor of 3.5. The interaction between the monolayer
    and the Au(111) surface (Au substrate effect) reduces the highest
    frequency gas phase mode by 276~cm$^{-1}$ and it increases its IR
    intensity by a factor of 2.6. The remaining third IR active gas phase mode
    is affected both by the Au(111) substrate and by the packing
    effect. See Sec.~\ref{sec:res_theory} for more detailed analysis.
    The calculated IR spectrum of an adamantane monolayer on Au(111) is
    shown with a red line, while the experimental spectrum is shown
    with a green line.}
\end{figure}

Our work combining IRSTM\cite{Ivan2013} measurements and \textit{ab
  initio} calculation of the IR spectrum of an
 adamantane monolayer on Au(111)
demonstrates the complex nature of {\it adamantane--adamantane} and
{\it adamantane--gold} interactions.  In Sec.~\ref{sec:res_theory} we
have described in detail the effect of each of these interactions on the
mixing (hybridization) of adamantane vibrational modes, the changes in
their frequencies, and the IR intensities. Figure~\ref{fig:summ}
summarizes our main results. The black dashed line in
Fig.~\ref{fig:summ} shows the calculated 
isolated adamantane gas phase IR spectrum,
while the red line shows the severely modified spectrum of the adamantane
monolayer on the Au(111) surface. The green line shows the experimental
spectrum of an adamantane submonolayer on Au(111).

The IR spectrum of the isolated adamantane molecule (black dashed line in
Fig.~\ref{fig:summ}) consists of three IR active C--H stretching modes
($\omega_5$, $\omega_6$, $\omega_7$). The {\it adamantane--adamantane}
interaction (packing effect) reduces the IR intensity of one of
these modes ($\omega_6$) by a factor of 3.5. On the other hand, the {\it
  adamantane--gold} interaction severely redshifts the gas phase
$\omega_5$ mode (by 276~cm$^{-1}$), and increases its IR intensity by
a factor of 2.6.  In addition, both $\omega_5$ and $\omega_6$ are hybridized
with IR inactive gas phase modes ($\omega_2$ and $\omega_1$
respectively). See Sections~\ref{sec:res_theory_monolayer},
\ref{sec:res_theory_gold} and Tables~\ref{tb:mono}, \ref{tb:gold}
for more details.

In conclusion, we expect that these techniques can be used to study
intermolecular and molecule--substrate effects of other molecular
systems, including the use of other metallic substrates. In particular, we
expect that the IR intensity reduction of the gas phase $\omega_6$
mode (or equivalent, for other molecules) can be used as a direct
measure of intermolecular interactions. Similarly, the increase in the
IR intensity and the redshift of the $\omega_5$ mode can be used as a
direct measure of molecule--substrate interactions.

\begin{acknowledgments}
  Computational resources were provided by the DOE at Lawrence
  Berkeley National Laboratory's NERSC facility.  Numerical
  calculations were also carried out on the TSUBAME2.0 supercomputer
  in the Tokyo Institute of Technology.  Theoretical part of the work
  was supported by NSF Grant No.~DMR-10-1006184
  (structural determination) and by the Nanomachines Program at
  the Lawrence Berkeley National Lab funded by the office of
  Basic Energy Sciences, DOE under Contract
  No.~DE-AC02-05CH11231 (infrared spectra simulations and analyses). 
  The Experimental part of the study was supported
  by the Nanomachines Program of the Office of Basic Energy Sciences,
  Materials Sciences and Engineering Division, U.S.~Department of
  Energy under Contract No.~DE-AC02-05CH11231 (STM measurements) and by
  the Department of Energy Early Career Award DE-SC0003949
  (development of IR laser source).  YS acknowledges financial support
  from Japan Society for the Promotion of Science.  RBC acknowledges
  financial support from Brazilian agencies CNPq, FAPERJ, INCT -
  Nanomateriais de Carbono and Rede de Pesquisa e Instrumenta\c c\~ao
  em Nano-Espectroscopia \'Optica. SGL acknowledges support of a
  Simons Foundation Fellowship in Theoretical Physics.
\end{acknowledgments}

\end{document}